\documentclass[12pt]{article}

\usepackage{array}
\usepackage{amsmath}
\usepackage{graphicx}

\def\bphi{\mbox{\boldmath $\phi$}}

\def\balpha{\mbox{\boldmath $\alpha$}}

\def\btau{\mbox{\boldmath $\tau$}}
\def\bnu{\mbox{\boldmath $\nu$}}
\def\bmu{\mbox{\boldmath $\mu$}}
\def\bpsi{\mbox{\boldmath $\psi$}}

\def\bt{\mbox{\bf t}}
\def\bc{\mbox{\bf c}}
\def\bu{\mbox{\bf u}}
\def\bb{\mbox{\bf b}}
\def\bs{\mbox{\bf s}}
\def\bd{\mbox{\bf d}}

\input{amssymb.sty}

\def\frac#1#2{{{#1}\over{#2}}}
\def\tfrac#1#2{{\textstyle{{#1}\over{#2}}}}

\def\half{\tfrac{1}{2}}

\begin{document}

\begin{titlepage}

\baselineskip 24pt

\begin{center}

{\Large {\bf A Comprehensive Mechanism Reproducing the Mass and Mixing
Parameters of Quarks and Leptons}}

\vspace{.5cm}

\baselineskip 14pt
  
  {\large Michael J BAKER and Jos\'e BORDES}\\
michael.baker@uv.es and jose.m.bordes\,@\,uv.es\\
{\it Departament Fisica Teorica and IFIC, Centro Mixto CSIC, Universitat de Valencia,
  Calle Dr. Moliner 50, E-46100 Burjassot (Valencia), Spain\footnote{Work supported by MEC and FEDER (EC) under grant FPA2011-23596 and Generalitat Valenciana under grant GVPROMETEO2010-056.}}\\

\vspace{.2cm}
{\large CHAN Hong-Mo}\\
h.m.chan\,@\,stfc.ac.uk \\
{\it Rutherford Appleton Laboratory,\\
  Chilton, Didcot, Oxon, OX11 0QX, United Kingdom}\\
\vspace{.2cm}
{\large TSOU Sheung Tsun}\\
tsou\,@\,maths.ox.ac.uk\\
{\it Mathematical Institute, University of Oxford,\\
  24-29 St. Giles', Oxford, OX1 3LB, United Kingdom}

\end{center}

\vspace{.3cm}

\begin{abstract}

It is shown that if, from the starting point of a universal
rank-one mass matrix long favoured by phenomenologists, one
adds the assumption that it rotates (changes its orientation 
in generation space) with changing scale, one can reproduce, 
in terms of only 6 real parameters, all the 16 mass ratios 
and mixing parameters of quarks and leptons.  Of these 16
quantities so reproduced, 10 for which data exist for direct 
comparison (i.e. the CKM elements including the CP-violating 
phase, the angles $\theta_{12}, \theta_{13}, \theta_{23}$ in 
$\nu$-oscillation, and the masses $m_c, m_\mu, m_e$) agree 
well with experiment, mostly to within experimental errors;
4 others ($m_s, m_u, m_d, m_{\nu_2}$), the experimental 
values for which can only be inferred, agree reasonably well;
while 2 others ($m_{\nu_1}, \delta_{CP}$ 
for leptons), not yet measured experimentally, remain as 
predictions.  In addition, one gets as bonuses, estimates 
for (i) the right-handed neutrino mass $m_{\nu_R}$ and (ii) the strong 
CP angle $\theta$ inherent in QCD.  One notes in particular 
that the output value for $\sin^2 2 \theta_{13}$ from the 
fit agrees very well with recent experiments.  By inputting 
the current experimental value with its error, one obtains 
further from the fit 2 new testable constraints: (i) that 
$\theta_{23}$ must depart from its ``maximal'' value: 
$\sin^2 2 \theta_{23} \sim 0.935 \pm 0.021$, (ii) that the 
CP-violating (Dirac) phase in the PMNS would be smaller than in the
CKM matrix: of order only $|\sin \delta_{CP}| \leq 0.31$ if not vanishing altogether.

\thispagestyle{empty}

\end{abstract} 

\end{titlepage}

\clearpage

\section{Introduction and Summary of Results}

The mass values and mixing matrices of quarks and leptons
together represent 20 independent quantities, accounting 
for some three-quarters of the total number of empirical 
parameters appearing in the otherwise very successful 
standard model.  They fall, besides, into a quite 
bewildering hierarchical pattern which cries out for a 
theoretical explanation.  But while their experimental 
colleagues have now measured some of these parameters to an 
impressive accuracy and are now giving shape even to the 
PMNS matrix so incredibly difficult to access, theoreticians, 
embarrassingly, have failed so far to produce a commonly 
accepted explanation even for their qualitative features, 
let alone their quantitative values to the accuracy that 
experiments have achieved.

Admittedly, the theoretician has no easy task here.  The 
explanation for these parameters cannot apparently be found 
in the standard model itself, since this logically seems to
admit any chosen values for these parameters.  On the other 
hand, most of the current models or theories which attempt 
to go beyond the standard model tend rather to increase the
number of independent parameters than to place restrictions 
on the existing ones.  There are models giving predictions
for some of the quantities, but not all.  For this reason, 
the mass and mixing pattern of quarks and leptons as a whole
still remains largely a mystery in particle physics which is
still to be understood.

The aim of this paper is to attempt a first understanding
of this great mystery via what has become known as the R2M2 
(rank-one rotating mass matrix) hypothesis.  This starts
from the assumption of a ``universal'' rank-one mass matrix
for all quarks and leptons long favoured by phenomenologists
\cite{Fritsch,Harari} as a first approximation and just adds 
to it the assumption that this matrix rotates (i.e. changes 
its orientation) with changing scale in generation space.  
A fermion mass matrix of rank one can always be rewritten by
a suitable relabelling of the $su(2)$-singlet right-handed
fields \cite{Weinberg} in the following form:
\begin{equation}
m = m_T {\balpha}{\balpha}^{\dagger},
\label{mfact}
\end{equation}
and by ``universal'' one means that the vector $\balpha$ 
is the same for all fermion species, that is,  whether up-type ($U$) 
or down-type ($D$) quarks, or whether charged leptons ($L$) or 
neutrinos ($N$).  The mass matrix (\ref{mfact}) has only one 
massive state for each fermion species, not a bad starting
point for the mass hierarchy, and the unit matrix for the mixing 
matrix (i.e. no mixing), also not a bad first approximation 
at least for quarks, hence the original phenomenological 
interest.  The rotation hypothesis in R2M2 is added to give
the needed deviations from this first approximation, such as
nonzero masses to the two lower generations and nontrivial
mixings as observed in experiment.  Notice that the fermion
mass matrix should rotate with changing scale occurs in the standard model \cite{Ramond}.  Thus, the only thing 
new in R2M2 is the assumption that it is the rotation which
gives rise to the mass hierarchy and to mixing rather than
the mixing which gives rise to the rotation as 
in the standard model.  Practically, this means that the mass
matrix has to rotate faster with respect to scale than it does
in the standard model. 

That the mass matrix rotates means that not only the masses
of the particles but also their state vectors in generation
space have to be defined each at its own mass scale.  That 
this will then lead immediately to nontrivial mixing and to 
nonzero masses for the two lower generations is easily seen 
as follows.  The state vector of the heaviest state in each 
fermion species is clearly to be taken as the vector $\balpha$ 
at its own mass scale.  Thus for the $t$ quark, the state 
vector is ${\bf t} = \balpha(\mu = m_t)$, and for the $b$ 
quark, ${\bf b} = \balpha(\mu = m_b)$.  The state vectors 
${\bf c}$ of $c$ and ${\bf u}$ of $u$ are then of necessity 
orthogonal to each other and both orthogonal to ${\bf t}$
forming together the $U$-triad.  Similarly, we have 
$({\bf b}, {\bf s}, {\bf d})$ the $D$-triad.  These two triads
are not aligned because the vector $\balpha$ by assumption
would have rotated from the scale $\mu = m_t$ to the scale
$\mu = m_b$, hence giving immediately nontrivial mixing.
Next, the vectors ${\bf c}$ and ${\bf u}$, being both by
definition orthogonal to ${\bf t} = \balpha(\mu = m_t)$, are
both null states of the mass matrix (\ref{mfact}) at $\mu =
m_t$.  But it is not to be inferred that $c$ and $u$ have
then zero masses for their masses have to be evaluated each 
at its own mass scale, not at $\mu = m_t$.  Thus, the mass 
$m_c$ has to be evaluated at $\mu = m_c < m_t$, at which
scale, the vector $\balpha$ would have rotated to a 
different direction and acquired a component in the ${\bf c}$
direction thus giving the $c$ quark a nonzero mass.  A similar
conclusion applies to the $u$ quark, and indeed to any quark 
or lepton of the two lower generations.  The nonzero masses 
that they thus acquire by this so-called ``leakage mechanism'' 
are small if the rotation is not too fast, hence giving rise 
to mass hierarchy, qualitatively as that seen in experiment.

A more detailed analysis in e.g. \cite{r2m2} along these 
lines yields the following formulae for the state vectors:
\begin{eqnarray}
{\bf t} & = & {\balpha}(m_t); \nonumber \\
{\bf c} & = & {\bf u} \times {\bf t}; \nonumber \\
{\bf u} & = & \frac{{\balpha}(m_c) \times {\balpha}(m_t)}
   {|{\balpha}(m_c) \times {\balpha}(m_t)|},
\label{tcutriad}
\end{eqnarray}
and for the masses:
\begin{eqnarray}
m_t & = & m_U, \nonumber \\
m_c & = & m_U |{\balpha}(m_c) \cdot{\bf c}|^2, \nonumber \\
m_u & = & m_U |{\balpha}(m_u) \cdot{\bf u}|^2.
\label{hiermass}
\end{eqnarray} 
Solving these equations together gives then the masses and
state vectors of each individual state.  Analogous results
hold for the $D$-type quark and leptons.  The mixing matrix
then follows for quarks as: 
\begin{equation}
V_{UD} = \left( \begin{array}{ccc}
   {\bf u} \cdot{\bf d}  &  {\bf u} \cdot
{\bf s}  &  {\bf u} \cdot{\bf b}  \\
    {\bf c} \cdot{\bf d}  &  {\bf c} \cdot
{\bf s}  &  {\bf c} \cdot{\bf b}  \\
    {\bf t} \cdot{\bf d}  &  {\bf t} \cdot
{\bf s}  &  {\bf t} \cdot{\bf b}  
          \end{array} \right),
\label{VUD}
\end{equation} 
with a similar mixing matrix $U_{LN}$ for leptons.

There are two points in the above result which need to be 
clarified, which the reader might have already
noted.  The first is that although the mass matrix, by 
assumption in R2M2, is of rank one throughout with two 
zero eigenvalues, and therefore invariant under chiral 
transformations in those eigenstates at every scale $\mu$, 
yet as a result of the rotation according to (\ref{hiermass}), 
all quarks and leptons have acquired a nonvanishing mass.
At first sight this may seem surprising, or perhaps even
counter-intuitive, it having been ingrained into us by 
experience working with nonrotating mass matrices, that 
the eigenvalues of a mass matrix are the mass values and
that chiral invariance is synonymous with the existence 
of zero mass states.  A rotating mass matrix, however, of
which not only the eigenvalues but also the eigenvectors 
are scale-dependent, is to most of us a new situation, and 
it stands to reason that for dealing with this we shall 
have to start afresh and check how many of those notions 
gained from experience before will still survive.  True, 
the mass matrix is already known to rotate in the standard 
model \cite{Ramond} and the question of how to deal with 
this situation should have arisen there.  However, 
the rotation there being slow, and its effects therefore
negligible, practically the question can be ignored, 
and has largely been ignored for this reason.  But it cannot now 
be ignored in R2M2.  Thus approaching the question anew 
with an open mind, it will become clear that for a mass 
matrix that rotates with scale, neither its eigenvalues 
nor its eigenvectors can in general be taken  
directly as the masses and state vectors of particles.  
Instead, the masses and state vectors are to appear as 
the eigenvalues and eigenvectors of a ``physical mass 
matrix'' to be constructed in a specific manner from the 
rotating matrix.  A detailed analysis of the situation
is what gives the result (\ref{tcutriad})---(\ref{VUD}) 
above \cite{r2m2}.  That being so, what might before have 
appeared as surprising or counter-intuitive will now no 
longer be the case.  
 
The second point to clarify concerns the Kobayashi-Maskawa
CP-violating phase in the mixing matrix.  Since the vector 
$\balpha$ is tacitly taken above to be real\footnote{There is in 
principle no need in R2M2 to assume that $\balpha$
is real.  Even if $\balpha$ is complex, however, the relative
phases between its 3 components will have to change with scale
$\mu$ before it can give rise to CP-violations in the CKM 
matrix, otherwise it is equivalent to a real $\balpha$.
We have taken $\balpha$ to be real here because (i) we can find
no empirical indication for the $\mu$-dependence of its phases,
(ii) we have not succeeded to build a model where $\balpha$ 
changes its phases with scale in the required manner, but most
importantly, (iii) we see here that CP-violations 
of sufficient size will in any 
case arise by virtue of the theta-angle term in QCD even for 
real $\balpha$.}, all the state 
vectors and the mixing matrix $V_{UD}$ in (\ref{VUD}) 
deduced from it will be real as well.  It might thus appear 
that R2M2 would not give us a proper CKM mixing matrix with 
a CP-violating phase.  Surprisingly, however, R2M2 contains
in itself a solution to this apparent problem, and in a most
intriguing manner, giving at the same time a solution to the
old strong CP problem in QCD \cite{Weinberg2}.  As is well
known, invariance principles allow in principle in the 
QCD action a CP-violating term of topological origin with
arbitrary coefficient $\theta$, which would give rise to, 
for example, an electric dipole moment for the neutron.  The
non-observance of this in experiment \cite{edm}
means, however, that this parameter $\theta$ has to be very
small, of order $10^{-9}$.  It is thus customary to insert
by fiat into the standard model an implicit assumption that
$\theta = 0$, or to add extra particles called axions \cite{axions}
which are not yet observed.
The need for doing so is known 
as the strong CP problem.  It is also well-known that the
theta-angle can be removed by a chiral transformation on 
the quark fields, but at the cost of making the quark mass
complex.  Hence, unless some quark happens to have zero mass
which is apparently at variance with experiment, the strong
CP problem still remains.  What is now interesting in R2M2, 
however, is the fact already noted in the above paragraph 
that the mass matrix can remain chirally invariant while
keeping all quark masses nonvanishing.  This means that the 
theta-angle term can be removed by a chiral transformation
without making the mass matrix complex, while keeping all
quark masses nonzero as experiment seems to demand.  There 
is, however, a penalty, namely that the chiral transformation
needed above to remove the theta-angle term is automatically
transmitted by the rotation to the quark mixing matrix and
makes it complex \cite{atof2cps}.  But this is, of course, 
a penalty one would be most willing to pay, for this phase
is exactly what one was missing in the mixing matrix $V_{UD}$
in (\ref{VUD}) above.  There is yet more, for it can readily 
be shown that provided that one starts with a $V_{UD}$ with
entries roughly similar in magnitude to those of the CKM 
matrix observed in experiment, then the CP-violation 
transmitted by rotation from a theta-angle of order unity 
will automatically be of the same order as that observed 
in experiment, i.e. corresponding to a Jarlskog invariant 
\cite{Jarlskog,Dalitz} $J$ of order $10^{-5}$.  

For details on the two points above and their clarification, 
the reader has to be referred elsewhere \cite{r2m2,atof2cps}.
Although the analysis is logically straightforward, it does 
take some care and patience to sort out, for which a freedom
from pre-conceived notions would seem, in our own experience, 
to be a prerequisite.  Some readers may find it convenient first tentatively to 
accept the R2M2 scheme as outlined above and see what 
it gives in practice before deciding to verify the details of 
its justification.  For this reason we first summarize the results before giving 
the details later in Section 2 and Appendix A of how they 
have been obtained.  Indeed, given the quality of the results, 
some readers may find the scheme a useful means of parametrizing the data, before getting into the theoretical details.

According then to the formulae (\ref{tcutriad})---(\ref{VUD}),
one should be able to evaluate the masses and mixing matrices
of both quarks and leptons once given $\balpha$ as a function
of the scale $\mu$.  As $\mu$ varies, the vector $\balpha$
traces out a curve on the unit sphere, which for our purpose is 
best described by the Serret--Frenet--Darboux formulae for curves 
lying on a surface, as will be spelt out in the next section.  
In this language then, the trajectory for $\balpha$ is 
completely prescribed first by giving the arc-length $s$ as a 
function of the scale $\mu$, and second by giving the geodesic 
curvature $\kappa_g$ of the traced-out curve, i.e. how much it 
is bending sideways, as a function of the arc-length $s$.  
Inspired by a model we have been studying, we choose to 
parametrize $s$ as an exponential in $\ln \mu$ (2 parameters) and 
$\kappa_g$ as some Breit--Wigner shape of $s$ (2 parameters).  
To these 4 real parameters, we have to add $\theta$, the 
theta-angle from the strong CP term to give the CKM matrix a 
CP-violating phase, as explained above.  Lastly, we add the Dirac mass $m_3^D$ of the heaviest neutrino $\nu_3$ as 
a parameter for in R2M2 the rotation mechanism depends on the 
Dirac masses, and these for neutrinos are obscured by the 
Majorana mass term of the right-handed neutrino(s) via a see-saw 
mechanism or something similar.  Altogether then, we have 6 
real parameters to our scheme, with $\theta$ expected to be 
of order unity.  

Once given a set of values for these parameters, the formulae
(\ref{tcutriad})---(\ref{VUD}) then allow one to calculate 
all the mixing matrix elements and masses of both quarks and 
leptons in terms of the coefficients $m_T$ in (\ref{mfact}),
one for each fermion species, which may be identified with 
the mass of the heaviest state in that species, and given the
values listed in Table \ref{M3}.  From these, one is then to 
reproduce the remaining 16 quantities, namely 4 from the CKM 
matrix, 4 from the PMNS matrix (neglecting for now any possible
Majorana phases), together with the 8 masses of 
the two lighter generations: $m_c, m_s, m_\mu, m_{\nu_2}, m_u, 
m_d, m_e, m_{\nu_1}$.

\begin{table}
\begin{eqnarray*}
\begin{array}{||c||c|c||}
\hline \hline
{\rm Quark/Lepton} & {\rm Mass \ used \ in \ fit \ (GeV)} & 
   {\rm Experiment} \\\hline \hline
t & 172.9 & 172.9 \pm 0.6 \pm 0.9 \\ \hline
b & 4.19 & 4.19^{+0.18}_{-0.06} \\ \hline
\tau & 1.777 & 1.77682 \pm 0.00016 \\ \hline
(\nu_3) & 0.020 & {\rm fitted \  parameter} \\ \hline \hline 
\end{array}
\end{eqnarray*}
\caption{Mass values for the heaviest generation used in the calculation
compared with data, where for the heaviest neutrino $\nu_3$ the entry
in the second column 
denotes the (unknown) Dirac mass.}
\label{M3}
\end{table}

By adjusting the values of the above 6 parameters, as detailed 
in the next section, one arrives without much difficulty at a 
set (\ref{params}) which gives the following results.  First, 
for the absolute values of the CKM matrix elements, one obtains 
the following:
\begin{equation}
|V_{\rm CKM}^{\rm th}| = \left( \begin{array}{lll}
 0.97423 & 0.2255 & 0.00414 \\
 0.2252 & 0.97342 & 0.0417 \\
 0.01237 & 0.0400 & 0.999122
\end{array}
\right),
\label{CKMth}
\end{equation}
and for the Jarlskog invariant, the value:
\begin{equation}
|J^q| = 2.91 \times 10^{-5}.
\label{Jqth}
\end{equation}
These are to be compared with experiment, for the
absolute values of the CKM matrix elements \cite{databook}:
\begin{equation}
\left( \begin{array}{ccc}
0.97428\pm0.00015&0.2253\pm0.0007&0.00347^{+0.00016}_{-0.00012}\\
0.2252\pm0.0007&0.97345^{+0.00015}_{-0.00016}&0.0410^{+0.0011}_{-0.0007}\\
0.00862^{+0.00026}_{-0.00020}&0.0403^{+0.0011}_{-0.0007}
&0.999152^{+0.000030}_{-0.000045}
\end{array} \right),
\label{CKMex}
\end{equation}
and for the Jarlskog invariant \cite{databook}:
\begin{equation}
|J^q| =\left( 2.91^{+0.19}_{-0.11}\right) \times 10^{-5}.
\label{Jex}
\end{equation}
One notes that apart from the two corner elements, $|V_{ub}|$ 
and $|V_{td}|$, all other quantities above are reproduced to 
within experimental errors.  The corner elements, being 
second order effects, as will be explained in the next section, 
point (a), are rather difficult to reproduce correctly by 
rotation using a simple parametrization such as equations
(\ref{sonmu}) and (\ref{kgons}),
but have nevertheless roughly the right size and the 
right asymmetry about the diagonals, as already anticipated 
\cite{cornerel}.

For the PMNS matrix, a first question to settle is whether
the chiral transformation applied on quarks above to remove 
the theta-angle term from the QCD action should affect leptons 
as well.  A priori, given that leptons carry no colour, they 
need have nothing to do with the theta-angle and have thus 
no reason to undergo a similar chiral transformation.  In that 
case, one would obtain for the PMNS matrix just the analogue 
$U_{LN}$ of (\ref{VUD}) above, which for the same values of 
our 6 parameters as before is found to take the form:
\begin{equation}
|U_{\rm PMNS}^{\rm th}| = \left( \begin{array}{rrr}
 0.818 & 0.556 & 0.147 \\
 0.519 & 0.605 & 0.604 \\
 0.247 & 0.570 & 0.784
\end{array}
\right),
\label{PMNSRth}
\end{equation}
with $J^\ell = \sin\delta_{CP}^\ell = 0$.  

The values of the $\nu$-oscillation angles $\sin^2 2\theta_{12}, 
\sin^2 2 \theta_{13}, \sin^2 2\theta_{23}$ corresponding to 
this PMNS matrix are listed together in Table \ref{nuos} and 
compared with experiment.  They are all seen to be within the 
experimental bounds given in \cite{databook}.

\begin{table}
\begin{eqnarray*}
\begin{array}{||c||c|c|c||}
\hline \hline
{\rm Source} & \sin^2 2 \theta_{12} & \sin^2 2 \theta_{23} & 
   \sin^2 2 \theta_{13} \\ \hline \hline
{\rm Our \ Output} & 0.864 & 0.935 &  0.0842 \\ \hline
{\rm Experiment} & 0.861^{+0.026}_{-0.022} & > 0.92 &  < 0.15 \\ \hline \hline 
\end{array}
\end{eqnarray*}
\caption{Output values of neutrino oscillation mixing angles 
compared with data \cite {databook}.}
\label{nuos}
\end{table} 

Since \cite{databook} was last updated, however, several major
experiments \cite{T2K,MINOS,DCHOOZ,DayaBay,RENO} have released 
new exciting results on $\theta_{13}$ which deserve a closer 
examination.  This will be done in section 3 where it will be 
seen that our output value as given in Table \ref{nuos} agrees 
very well with the new data (as summarized in Tables 4 and 5).

That there is no need for a chiral transformation on leptons
to remove the theta-angle term, however, does not necessarily
mean that one cannot be applied nevertheless to give the PMNS 
matrix a CP-violating phase.  This possibility will be examined 
in Appendix B where it will be shown that in view of the recent 
data on $\theta_{13}$, the present R2M2 fit can admit only a 
small value for $\sin \delta^\ell_{CP}$ arising in this way.  In 
addition, considerations will be given there to a situation 
when such a CP-violating phase may become a theoretical necessity.  

Lastly, the quark and lepton masses we obtained are listed
together and compared with experiment in Table \ref{massout}.
The agreement is very good for $c$ and $\mu$.  It is not bad 
even for $s, e$ and $\nu_2$ where some delicate extrapolation
is involved.  For the others $u, d$ and $\nu_1$, no ready 
information is available yet for a meaningful comparison. 
A more detailed discussion on the last two points will  
be given in the next section.  
\begin{table}
\begin{eqnarray*}
\begin{array}{||c||l|l||}
\hline \hline
{\rm Quark/Lepton} & {\rm Output\ mass\ in\ GeV} & {\rm Experiment} \\
\hline \hline
c & 1.234 & 1.29^{+0.05}_{-0.11} \\ 
s & 0.171^* & {0.100^*}^{+0.030}_{-0.020} \\ 
\mu & 0.10564 & 0.10565837 \\
\nu_2 & 0.0064^{\dagger} & 0.0083^{\dagger} \\ \hline
u & 0.0009^* & 0.0017 - 0.0033 \\
d & 0.0008^* & 0.0041 - 0.0058 \\
e & 0.0007 & 0.0005109989 \\
\nu_1 & < 0.0016^{\dagger} & {\rm unknown} \\ \hline \hline
\end{array}
\end{eqnarray*}
\caption{Output masses compared with data.  Note that those
entries marked by a * or a $\dagger$ cannot be compared with
data directly.  For the light quarks $u, d, s$ marked with *
the output values are supposed to represent the mass of the
free quark measured at its own mass scale while values given
in the PDG review \cite{databook}, because of quark confinement,
are determined at a scale of 2 GeV.  For the neutrinos, marked
with $\dagger$, the output masses are the Dirac masses which
are not measured directly.  The ``experimental values'' cited
are inferred from the measured (physical) mass differences via 
a see-saw mechanism.  For more details, see text, Section 2.}
\label{massout}
\end{table} 

In summary, of the 16 quantities reproduced, 10 of which can be 
compared directly with experiment namely: 4 for the CKM matrix, 
the 3 mixing angles $\theta_{12}, \theta_{23}, \theta_{13}$ 
from neutrino oscillation, and the masses $m_c, m_\mu, m_e$, 
agree with experiment, mostly to within present experimental 
errors, or else to a high accuracy in the case of $m_\mu$.  
The only exceptions are the corner CKM elements $V_{ub},V_{td}$ 
and the electron mass $m_e$, the agreement for which is only
approximate and outside experimental errors, for reasons that 
will be given later in the next section, points (a) and (b).  Of 
the remaining 6, 4 can be compared only with values inferred from 
experiment, where $m_s, m_{\nu_2}$ agree within reason but $m_u, 
m_d$ only in the rough order of magnitude, again for reasons to 
be given in the next section, points (d) and (e).  This leaves 2 
others $m_{\nu_1}$ and $\delta^\ell_{CP}$ for leptons as predictions, 
for which no experimental information is yet available.  In 
addition, one obtains as bonuses estimates for the right-handed 
neutrino mass $m_{\nu_R}$ and the strong CP angle $\theta$, for 
which no experimental information is likely to be available for 
some time, and therefore for the moment of theoretical interest
only.  Although $\theta$ was treated above as an input parameter 
and used to fit the value of the Jarlskog invariant $J^q$ for the 
CKM matrix, that this was possible is only by virtue of R2M2 so
that a value for $\theta$ should itself be counted as an output.  
Altogether then, one has obtained 18 quantities for the price of 
6, which is not too bad a bargain.  What is perhaps more relevant, 
however, is that one has obtained, probably for the first time 
ever, a comprehensive picture of the whole mass and mixing 
pattern of both quarks and leptons with a single all-embracing 
mechanism and the same parameters throughout.

\begin{figure}
\includegraphics[height=13.5cm]{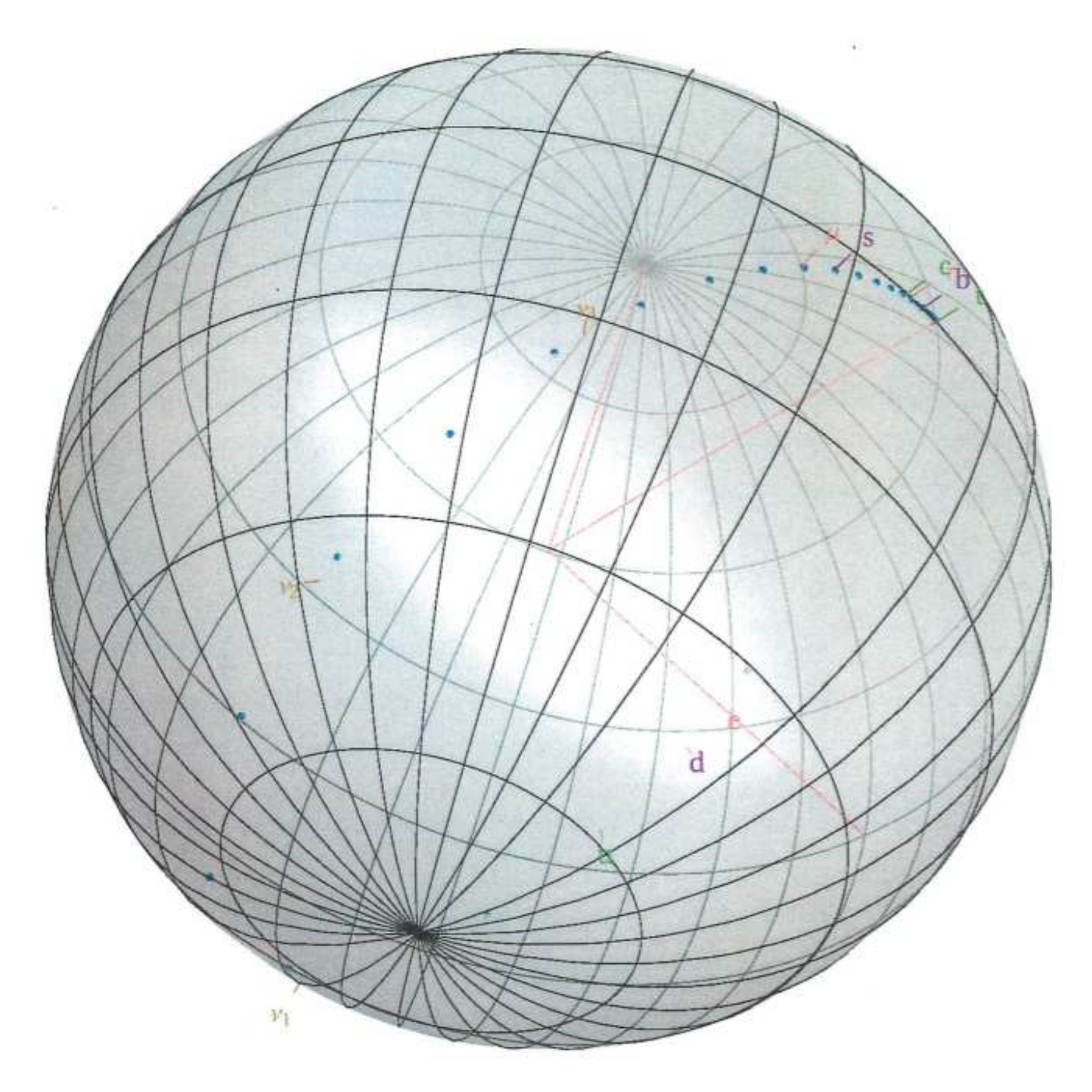}
\caption{The trajectory traced out by $\balpha$ on the unit 
sphere as the scale $\mu$ decreases over some 6 decades in 
orders of magnitude from $\mu \sim m_t$ to $\mu \sim m_e$.  
The points shown on the curve are spaced at equal intervals of 
$1/5$ of a decade in $\ln \mu$ indicating 
from their spacing the ``local'' speed of rotation at any $\mu$.  
The rotation is seen to accelerate as $\mu$ decreases, while 
the geodesic curvature (i.e. sideways bending) of the curve 
traced out is reduced.  Also, along the trajectory, are marked 
out the individual mass scales of all the 12 quark and lepton 
states to show the position they each occupy, where we note 
that $u, d$ and $e$ are located at the back of the sphere.}  
\label{Muisphere}
\end{figure}

The trajectory for $\balpha$ from which these results are obtained 
is shown in Figure \ref{Muisphere}.  It is seen to possess certain 
characteristics, e.g. that it accelerates in rotation as the scale 
$\mu$ decreases, while the (geodesic) curvature of the curve it 
traces out is reduced, in a manner to be detailed in the next 
section, but is otherwise quite innocuous.  

More details of the calculation will be supplied in Appendix 
A so that, with the help of Figure \ref{Muisphere}, interested 
readers can readily verify for themselves the validity of the 
assertions made as well as the numerical values of the output 
quantities cited.

\section{The Mechanism and How it Works}

As the scale changes, the vector $\balpha$ traces out a curve 
on the unit sphere, for which it is found convenient to 
adopt the Serret--Frenet--Darboux formalism for curves on a 
surface \cite{Docarmo}.  At every point on the curve, one 
sets up then a Darboux triad, consisting of first the normal 
${\bf n}$ to the surface, then the tangent vector $\btau$ to 
the curve, and thirdly the normal $\bnu$ to both the above, 
all normalized and orthogonal to one another.  The 
Serret--Frenet--Darboux formulae then say:
\begin{eqnarray}
{\bf n}'={\bf n}(s+\delta s) & =& {\bf n}(s) - \kappa_n \btau (s)\,\delta s 
+ \tau_g \bnu(s)\,\delta s,  \nonumber \\
\btau'=\btau(s+\delta s) &=& \btau (s) + \kappa_n{\bf n}(s)\,\delta s
+ \kappa_g \bnu (s)\,\delta s,  \nonumber \\
\bnu' = \bnu(s+\delta s) &=& \bnu(s) - \kappa_g \btau (s)\,\delta s -\tau_g
{\bf n}(s)\,\delta s,
\label{SFD1}
\end{eqnarray}
correct to first order in $\delta s$, a small increment in 
the arc-length $s$, where $\kappa_n$ is known as the normal 
curvature (bending along the surface), $\kappa_g$ the geodesic 
curvature (bending sideways) and $\tau_g$ the geodesic torsion 
(twist).  For the special case here of a curve on the unit 
sphere, the surface normal ${\bf n}$ is the radial vector 
$\balpha$, $\kappa_n = 1$ and $\tau_g = 0$, so that the 
formulae reduce to:
\begin{eqnarray}
\balpha'=\balpha(s+\delta s) & =& \balpha(s) - \btau (s)\,\delta s, 
\nonumber \\
\btau'=\btau(s+\delta s) &=& \btau (s) + \balpha(s)\,\delta s
+ \kappa_g \bnu (s)\,\delta s,  \nonumber \\
\bnu' = \bnu(s+\delta s) &=& \bnu(s) - \kappa_g \btau (s)\,\delta s.
\label{SFD2}
\end{eqnarray}
The curve will thus be specified just by giving the geodesic
curvature $\kappa_g$ as a function of the arc-length $s$.

For the problem here, one is interested not only in the shape
of the curve traced out by $\balpha$ but also in the speed with 
respect to the scale $\mu$ at which $\balpha$ moves along that 
curve.  To specify the ``trajectory'' of $\balpha$, then, one 
will need to give not only $\kappa_g$ as a function of $s$ but 
also $s$ as a function of $\mu$.

These two functions we propose now to parametrize as follows,
inspired by a dynamical model \cite{prepsm,efgt,dfsm} we have 
been considering.  For $s$ as a function of $\ln(\mu)$ ($\mu$ in GeV) we propose 
an exponential, thus:
\begin{equation}
s = \exp[-a-b\ln(\mu)] - \exp[-a-b\ln(m_t)]
\label{sonmu}
\end{equation}
and for $\kappa_g$ as a function of $s$, we propose the tail 
of a square-root Breit-Wigner function, namely:
\begin{equation}
\kappa_g = \frac{g}{\sqrt{(s - s_0)^2 + s_0^2}},
\label{kgons}
\end{equation}
with $g$ being the residue, and $s_0$ denoting the
position of a pole near the physical region.  The reason
for these proposals is that in the model cited, there is a
fixed point for $\balpha$ near $\mu = m_t$, where the speed at which $\balpha$ rotates tends to zero
while the geodesic curvature $\kappa_g$ tends to infinity 
\cite{dfsm}.  What seems to
matter is only the distance of $t$ to the pole, not the ratio of
its real and imaginary parts,
so that we can use just the one parameter $s_0$ throughout.  However, there is nothing very special about 
these parametrizations and any similar choice of functions 
would likely lead to similar results.

With our choice of (\ref{sonmu}) and of (\ref{kgons}) above, 
the trajectory for $\balpha$ is specified by the 4 parameters
$a, b, g, s_0$.  Once given a set
of values for the 4 parameters $a, b, g, s_0$, one can
reconstruct the trajectory by integrating the equations in
(\ref{SFD2}).  And once one has the trajectory, one can then
calculate the mass ratios and the mixing parameters for both 
quarks and leptons. 

To calculate the masses of the members of any fermion species, 
say the $U$-type quarks $t, c, u$, one has first to normalize 
by supplying, say the mass $m_U = m_t$ of the heaviest state 
$t$ as listed in Table \ref{M3}.  The mass of the $c$ quark 
is then obtained as per (\ref{tcutriad}) and (\ref{hiermass})
by solving the equation:
\begin{equation}
m_c = m_U |{\bf c}\cdot\balpha(\mu = m_c)|^2.
\label{mc}
\end{equation}
And once we know the masses of $t$ and $c$, we can then solve 
(\ref{tcutriad}) for all the state vectors $\bt, \bc, \bu$ of
the $U$-triad.  A similar procedure applies for the $D$-type
quarks to give the $D$-triad.  Combining the results, we can
then calculate the matrix $V_{UD}$ in (\ref{VUD}) above.

As noted before, however, this real mixing matrix $V_{UD}$ is 
not yet the CKM matrix we seek.  If there is a nonvanishing 
theta-angle term in the QCD action, as there is in general no
reason not to have in theory, then strong interaction will be
CP-violating, which is a situation experiment will find hard 
to accommodate.  However, as was mentioned above and worked out in 
detail in \cite{atof2cps}, this unwanted theta-angle term can 
be eliminated in the present R2M2 scheme, hence making the 
strong sector CP-conserving as desired, by a judicious chiral 
transformation on the quark state vectors, thus making them 
complex.  The easiest way to specify the effect of this is as 
follows \cite{r2m2}.  At $\mu = m_t$, we recall that $\balpha$ 
coincides with the state vector ${\bf t}$ for the $t$ quark.  
The state vectors ${\bf c}$ and ${\bf u}$ must then be 
orthogonal to $\balpha$ there and to each other, the three 
quarks $t$, $c$ and $u$ being by definition independent 
quantum states.  They must therefore be related to $\btau$ and 
$\bnu$, the other two members of the Darboux triad at $\mu = 
m_t$, which are also orthogonal to $\balpha$, just by a 
rotation about $\balpha$, say by an angle $\omega_U$, thus:
\begin{equation}
({\bf t}, {\bf c}, {\bf u}) = (\balpha, \cos \omega_U \btau
+ \sin \omega_U \bnu, \cos \omega_U \bnu - \sin \omega_U \btau).
\label{Utriad}
\end{equation}
Similarly, for the $D$-type quarks, we have:
\begin{equation}
({\bf b}, {\bf s}, {\bf d}) = (\balpha', \cos \omega_D \btau'
+ \sin \omega_D \bnu', \cos \omega_D \bnu' - \sin \omega_D \btau'),
\label{Dtriad}
\end{equation}
where $(\balpha', \btau', \bnu')$ is now the Darboux triad 
taken at $\mu = m_b$.  What the chiral transformation which 
eliminates the theta-angle term does is to give the $\bnu$ 
component of each of the above vectors a phase $\exp (-i 
\theta/2)$, these becoming thus:
\begin{align}
(\tilde{\bt},\tilde{\bc},\tilde{\bu}) &= (\balpha,\,\cos \omega_U \btau +\sin 
\omega_U \bnu\,
e^{-i \theta/2},   \,\cos \omega_U \bnu\, e^{-i\theta/2}
\!-\sin \omega_U \btau) \\
(\tilde{\bb},\tilde{\bs},\tilde{\bd})& = (\balpha', \,\cos \omega_D \btau' + 
\sin \omega_D
\bnu'\, e^{-i \theta/2},  \,\cos \omega_D \bnu' \,e^{-i  \theta/2}
\!-\sin \omega_D \btau').
\label{UDtriadsCP} 
\end{align}
The CKM matrix can then be evaluated with these two new triads 
of state vectors as before by (\ref{VUD}), but this will now 
be the true complex CKM matrix containing a CP-violating phase.  
For instance, if one calculates the Jarlskog invariant\footnote{\label{Jarlfn}Note however
 that in R2M2, where the orientation of the mass matrix is scale
 dependent, CP-violation in the CKM matrix cannot be directly
 related to commutators of (hermitian) mass matrices as proposed
 by Jarlskog in \cite{Jarlskog}.  Instead, one has to rely solely
 on the unitarity properties of the mixing matrix and work with
 the quartic rephasing invariants which are scale independent.
 These invariants appeared in earlier work on CP-nonconservation
 \cite{Dalitz} without use being made of the mass matrix commutators.} of this 
new matrix, one will obtain a nonzero answer.  On the other hand, 
the formulae for the masses of the various quark states are not 
affected by this chiral transformation \cite{r2m2}.

The same procedure as the above for calculating the quark masses 
applies to the charged lepton states.  For the neutrinos, there 
is a difference in that the R2M2 mechanism applies to the 
Dirac masses, which for neutrinos are unknown, being obscured via 
a possible see-saw mechanism \cite{seesaw} by the Majorana mass 
term for the right-handed neutrino.  We have therefore to supply the Dirac 
mass $m_{\nu_3}^D$ of the heaviest neutrino $\nu_3$ as a parameter, 
as listed in Table \ref{M3} above.  Once given some value for 
$m_{\nu_3}^D$, the same procedure as before will yield the Dirac 
masses of the 2 lighter neutrinos.  To compare these output values 
with experiment, the Dirac masses have to be inferred from the 
physical masses of the neutrinos measured in experiment by 
assuming some specific see-saw mechanism.  Here, we have taken 
the simplest model: Type I quadratic see-saw mechanism 
\cite{seesaw} in which the physical masses of the neutrinos 
are given as:
\begin{equation}
m_{\nu_i} = (m_{\nu_i}^D)^2/m_{\nu_R}.
\label{numass}
\end{equation}
It is the values inferred in this way which are entered in the
Table \ref{massout} for comparison with the output values, and
gives incidentally also an estimate for the right-handed neutrino mass
$m_{\nu_R}$.

The same procedure as the above for calculating the CKM matrix 
applies in principle also to the PMNS matrix except for the 
question whether the chiral transformation for eliminating the 
theta-angle term in the QCD action should affect the leptons 
as well.  Here, in the text, as was explained, we shall consider 
only the case when it does not, leaving the other case to be 
considered in Appendix B.  The PMNS matrix for leptons is then 
real and can be calculated in the same way as the $UD$-matrix for 
quarks in (\ref{VUD}) above.  As already mentioned, we neglect for now
any possible Majorana phases, the detection of which appears to be
only possible in the remote future \cite{majoranaphase}.

Having now completed the specification of how the 16 mass 
ratios and mixing parameters are calculated given any set
of values for the 6 parameters $a, b, g, s_0, \theta, 
m_{\nu_3}^D$, we can now start varying the input values of the 
6 parameters to see whether any choice of those can give a 
decent description of the data.  Before doing so, it is worth 
gaining first some qualitative understanding of how the value 
of each parameter will affect the agreement of the output with 
experiment.

Let us start with the parameters $a$ and $b$, which govern the
dependence of $s$ on $\mu$.  Some quantities, such as the mass
of the muon $m_\mu$ and the CKM element $V_{tb}$ depend rather
simply each on the rotation angle between two states, namely
between $\btau$ and $\bmu$ for $m_\mu$, and between $\bt$ and
$\bb$ for $V_{tb}$.  They would thus be sensitive to how the
arc-length $s$, or the angle subtended at the centre of the
sphere, depends on the scale $\mu$, or in other words to the
parameters $a$ and $b$.  The two quantities $m_\mu, V_{tb}$ 
have also both been determined accurately by experiment, 
and so can be used as anchors for these two parameters.  Indeed, 
based more or less just on this observation, approximate values 
for $a, b$ have already been obtained some 10 years ago 
\cite{cevidsm}, and the new values we shall suggest will not 
deviate much from those.

Next, for the two parameters $g$ and $s_0$ governing
the dependence of $\kappa_g$ on $s$, it is instructive first to 
examine the situation when $\theta_{tb}$, the angle between 
$\bt = \balpha(\mu = m_t)$ and $\bb = \balpha(\mu = m_b)$, is
small, which indeed it is experimentally, its value estimated 
from the experimental value of $V_{tb}$ by the consideration 
in the above paragraph being only $\sim 0.041$.  Substituting 
$\theta_{tb}$ for $\delta s$ in (\ref{SFD2}),(\ref{Utriad}),
(\ref{Dtriad}) then easily yields the following approximate 
formula for the mixing matrix (\ref{VUD}) \cite{cornerel}:
\begin{equation}
V_{UD} = \left( \begin{array}{ccc} \cos \omega - \kappa_g \sin \omega
    \,\theta_{tb} &  \sin \omega + \kappa_g \cos \omega \,\theta_{tb} & 
     \sin \omega_U \,\theta_{tb} \\ -\sin \omega - \kappa_g \cos \omega 
    \,\theta_{tb} & \cos \omega - \kappa_g \sin \omega \,\theta_{tb} &
    -\cos \omega_U \,\theta_{tb} \\ -\sin \omega_D \,\theta_{tb} & 
    \cos \omega_D \,\theta_{tb} & 1 \end{array} \right)
\label{Vckm2}
\end{equation}
with $\omega = \omega_D - \omega_U$.  We notice there that 
the geodesic curvature $\kappa_g$ appears prominently in the
Cabibbo angle $V_{us}, V_{cd}$.  Again, the Cabibbo angle,
being well measured experimentally, can be used as an anchor
for the parameters $g$ and $s_0$ appearing in (\ref{kgons}).  
The same observation applies qualitatively to the solar neutrino 
angle $U_{e2}$ in the leptonic equivalent matrix $U_{LN}$, 
although the angles involved in this case are no longer small.  
Thus, by adjusting the parameter $g$, one can fit roughly the 
output value of the Cabibbo angle to its experimental value, 
and by adjusting the position of the pole, namely $s_0$ 
in (\ref{kgons}), the relative magnitude of $\kappa_g$ between 
the quark and lepton sectors and hence the relative magnitude 
of $V_{us}$ to $U_{e2}$.  

There are two things still missing in the preceding arguments,
namely the effect of the theta-angle on the phases of the
mixing matrices, and that of the Dirac mass of $\nu_3$.  The
former affects mostly the Jarlskog invariant $J^q$ for quarks, 
while the latter, by specifying where the state vector $\bnu_3$ 
will lie on the trajectory, mostly the atmospheric angle 
$U_{\mu 3}$, and can each be used to anchor the corresponding 
quantity.  However, they will both also affect the quantities 
already considered, as will also the other 4 parameters 
affect the present 2 quantities, though all to a smaller extent.  
Nevertheless, armed with these preliminary understanding of 
how the values of our 6 parameters will affect mainly which 
of the output quantities, it is not difficult to jiggle them 
around to achieve a tolerably decent fit of the quantities 
concerned.  Working in this way, one arrives at the following 
set of values for our parameters, which is what gives the 
result summarized in the preceding section:
\begin{equation}
a  = 2.27635; \ b = 0.512; \ g = 0.42214; \ s_0= -0.069;
  \ \theta = 1.28; \ m_{\nu_3}^D = 20\;{\rm MeV}.
\label{params}
\end{equation}

There is of course nothing immutable about these parameters,
nor indeed even about the parametrizations (\ref{sonmu}) 
and (\ref{kgons}).  They are meant to represent just a sample 
fit by R2M2 to data, and any parametrization which gives a 
trajectory for $\balpha$ with a similar shape to that shown 
in Figure \ref{Muisphere} is likely to do quite as well.  For
instance, there has been no serious attempt to optimize the
parameters in (\ref{params}) on our part, for indeed, given
the great disparity in accuracy achieved in experiment for the 
16 quantities we are after, it is not clear what optimization
of the parameters should mean in fitting them.  What we wish
to assert is that with the above sample choice, one seems able to obtain an adequate description of the data.

Notice that with the 6 parameters we have permitted ourselves,
we can anchor but 6 of our output quantities by experiment, and
these 6 quantities, as seen, can all be fitted without trouble
to within experimental errors or to a high accuracy as in the
case of the $\mu$ mass.  But the other 10 quantities we also 
want will then have to fend for themselves.  That they all have 
managed to do so reasonably well should thus be regarded as a 
test of the R2M2 hypothesis.

Before moving on further, the following notes on some points 
of detail would be in order:

(a) First, one notes that the corner elements in (\ref{Vckm2}) 
are proportional to $\sin\omega_{U,D}\, \theta_{tb}$, where each of 
the angles is itself a consequence of the rotation.  These 
corner elements $V_{ub}$ and $V_{td}$, therefore, are  
second order effects of the rotation \cite{cornerel}.  This
comes about in R2M2 just from the geometrical fact that for a
curve on the unit sphere, the geodesic torsion vanishes so that
any twist that these elements represent will have to be of at
least the second order in the change in arc length $\delta s$. 
It thus follows immediately that they are small, as experiment
shows.  It also follows that they have the asymmetry $V_{td}
> V_{ub}$, again as experiment wants, since by virtue of the 
behaviour (\ref{sonmu}) of $s$ as a function of $\mu$, one has
$\omega_D > \omega_U$.  For more details on these points, see 
\cite{cornerel}.  However, the down side is that, being 
second order effects in rotation, these corner elements are 
rather difficult to reproduce accurately in the R2M2 scheme, 
hence their output values in (\ref{CKMth}) lying outside the 
experimental errors in (\ref{CKMex}) above. 

(b) Secondly, one notes that of the parameters in (\ref{params}), 
4 ($a, b, g, \theta$) are anchored to experiment in the high
scale region with $\mu > 100$ MeV, using data there of 
high accuracy, namely $m_\mu, V_{tb}, V_{us} \sim V_{cd}$ 
and $J^q$ for quarks.  The other 2, namely $s_0$ and 
$m^D_{\nu_3}$, are anchored on neutrino oscillation data of 
understandably lower accuracy.  This means that the rotation 
trajectory is quite well determined in the region $\mu > 100$ MeV, less well but still adequately down to about 
$m^D_{\nu_2} \sim 6$ MeV, below which, however, it is 
obtained just by extrapolation from the high scale region. 
This extrapolation does not affect, of course, the masses of 
the two heavier generations, nor the two mixing matrices
since these depend only on the state vectors which, including
those of the lightest generation, are already determined by 
(\ref{tcutriad}) at the scale of the middle generation.  The 
masses of the lightest generation, however, depend on the 
extrapolation, and given the assumed exponential shape of 
(\ref{sonmu}), this can cover quite a considerable range in 
arc-length $s$, as is seen in Figure \ref{Muisphere}.  In view 
of this, that one is able to get a roughly correct value (i.e. 
0.7 as compared to 0.5 MeV) for the $e$ mass is already better 
than can reasonably be expected.  In any case, reproducing the 
mass of the electron being such an audacious undertaking in 
itself, an error of this sort probably needs no apology.  

(c) Thirdly, again with reference to the $e$ mass, and those 
of $u$ and $d$ in the lightest generation, one notes that 
there are in fact more than one solution each to the equations 
(\ref{hiermass}).  We recall that at the mass scale of the 
middle generation, say $c$ to be specific, the state ${\bf u}$ 
has zero eigenvalue, but as the scale decreases from $m_c$, 
$\balpha(\mu)$ will acquire a component $\langle {\bf u}|
\balpha(\mu) \rangle$ in the direction of ${\bf u}$.  This will 
gradually increase in size until at some point $\langle {\bf u}|
\balpha(\mu) \rangle$ matches the value $\sqrt{\mu/m_t}$ and we 
have, according to (\ref{hiermass}), a solution, say the first 
(higher) solution, to the $u$ mass.  However, as $\mu$ decreases 
further, $\langle {\bf u}|\balpha(\mu) \rangle$ may vanish again, 
in which case one may find in the neighbourhood again a solution 
for $\langle {\bf u}|\balpha(\mu) \rangle = \sqrt{\mu/m_t}$, 
namely a second (lower) solution to (\ref{hiermass}) for the
$u$ mass.  Indeed, this is what happens to the trajectory here
and the mass values cited in Table \ref{massout} for $u$, 
$d$, and $e$ correspond to this lower, likely more stable solution.  
The limit cited in Table \ref{massout} for the Dirac mass of 
the lightest neutrino $\nu_1$, however, corresponds to the 
higher solution since the lower solution for $m^D_{\nu_1}$ has 
not yet been found in the range of scales explored and we are 
reluctant to trust the extrapolation further.  An interesting 
phenomenological point to note here is that given the great 
disparity in value between the normalization factors $m_T$ in 
(\ref{mfact}) for the 3 species $U, D, L$, namely the masses 
$m_t, m_b, m_\tau$, one would have 
expected a similar disparity between the output masses of the 
lower generations obtained by leakage from the heaviest.  This 
is indeed borne out for the middle generation, as seen in Table
\ref{massout}, as well as for the higher solution to $u$, $d$
and $e$ (not shown), which latter would be in stark 
contradiction to experiment.  However, as seen in Table 
\ref{massout}, one gets from the lower solution similar masses 
of order MeV for all 3 species $u, d, e$, which is apparently 
what experiment wants.  The question why the lightest generation, 
in particular, is prone to having such lower solutions to their 
mass in the R2M2 scheme is an interesting one deserving more
clarification, which will be supplied later, with other details,
in Appendix C.

(d) Fourthly, quarks being confined, experiment cannot measure 
the masses of the lighter quarks $s, u, d$ directly at their 
own mass scales but only at a higher scale, usually taken to be
around 1 or 2 GeV.  Since the perturbative method currently 
available for QCD does not allow for running to such low scales,
there is no way at present to compare the output masses of 
the light quarks determined at their own mass scales with their 
masses given in experiment, e.g. in \cite{databook}.  For the 
$s$ quark, the mass scale being of order $\sim 100\ {\rm MeV}$, 
not that much below the scale 1--2 GeV at which it is measured, 
one can perhaps claim that our output value of 170 MeV at its 
own mass scale is in reasonable agreement with the value of 
about 100 MeV given by experiment at a scale of 1--2 GeV.  For 
the $u$ and $d$ quarks, however, one can go no further than 
saying that the output values are roughly of the same MeV 
order as the experimentally given values.  We have thus no 
explanation whatsoever so far of the crucial fact why the $u$ quark 
should be lighter than the $d$.

(e) Lastly, based on the see-saw mechanism outlined above, one
can estimate also the physical masses of the 3 generations of 
neutrinos, as respectively $0.05, 0.005, < 0.0003$ eV, 
normalized at $m_{\nu_3} \sim 0.05$ eV, as estimated 
from $\delta m^2_{23} \sim  2.43 \times 10^{-3}$ ${\rm eV}^2$. 
Here we have normal hierarchy,  
as is natural from the leakage mechanism and universality that R2M2 
implies.  Notice that the above output values for the physical 
masses of $\nu_2$ and $\nu_1$ would correspond to a value of 
$\delta m^2_{12}$ of only $\sim 2.5 \times 10^{-5}$ ${\rm eV}^2$, 
i.e., a factor of 3 smaller than the experimental value of 
$7.59 \pm 0.21 \times 10^{-5}$ ${\rm eV}^2$ \cite{databook}.  
However, since $\delta m^2_{12}$ is proportional to the 8th 
power of the rotation angle calculated by the R2M2 scheme, this 
factor of 3 translates only to an error of about 15 percent in 
the rotation angle, which is not bad in view of the extrapolation 
involved as detailed in (b).  These masses correspond 
to a right-handed neutrino mass of about 8000 TeV, 
and, as far we know, there are
no stringent empirical bounds on this \cite{rhnu}.

\section{A Closer Look at Lepton Mixing and the PMNS Matrix}

Since the PDG website \cite{databook} giving the bounds quoted in Table
\ref{nuos} was last updated, 5 experiments \cite{T2K,MINOS,
DCHOOZ,DayaBay, RENO} have given exciting new results on the 
mixing angle $\theta_{13}$ which now need to be taken into 
account.  We shall do so in the order in which these results
appeared. 
  
As regards the two appearance experiments T2K and MINOS, the
quantity they quote is $2 \sin^2 \theta_{23}\sin^2 2 \theta_{13}$, and the result depends on the CP phase
$\delta^\ell_{CP}$.  It is thus the quoted value of this quantity
at $\delta^\ell_{CP} = 0$ that should be compared with the value
of $2 \sin^2 \theta_{23} \sin^2 2 \theta_{13}$ from our output
above in Table \ref{nuos}.  The result of this comparison is 
given in Table \ref{cfT2KMINOS} where it is seen that our 
output value agrees with the experiments to the extent that the two 
experiments agree between themselves.

\begin{table}
$$
\begin{array}{||c||c|c||}
\hline \hline
\multicolumn{2}{||c||}{\rm Source} &  2 \sin^2 \theta_{23}
\sin^2 2 \theta_{13} \\ \hline \hline
{\rm Our \ Output} & & 0.0628 \\ \hline
{\rm T2K} & 68\% {\rm C.L.} & 0.12^{+0.09}_{-0.06} \\ \cline{2-3}
          & 90\% {\rm C.L.} & 0.12^{+0.16}_{-0.07} \\ \hline
{\rm MINOS} & 68\% {\rm C.L.} & 0.04^{+0.05}_{-0.03}  \\ \cline{2-3}
          & 90\% {\rm C.L.} & 0.04^{+0.08}_{-0.04}  \\ \hline \hline
\end{array}
$$
\caption{Output compared with the values of $2 \sin^2 
\theta_{23} \sin^2 2 \theta_{13}$ at $\delta^\ell_{CP} = 0$ read 
from \cite{T2K,MINOS}.}
\label{cfT2KMINOS}
\end{table} 

The other three disappearance experiments, Double Chooz, Daya 
Bay, and RENO measured simply $\sin^2 2 \theta_{13}$ and the 
result does not depend on the CP phase $\delta^\ell_{CP}$ so 
their result can be directly compared with our output value, 
as is done in Table \ref{cftherest}.  The agreement is seen 
to be very good.

Finally one can compare our output value with the combined fit 
for $\sin^2 2 \theta_{13}$ of the results from the first four 
experiments given by A. Blondel in his summary talk at the 
recent Moriond meeting \cite{Blondel}.  As seen also in Table 
\ref{cftherest}, our output value falls right in the middle of 
the allowed experimental range.

\begin{table}
\begin{eqnarray*}
\begin{array}{||c||c||}
\hline \hline
{\rm Source} & \sin^2 2 \theta_{13} \\ \hline \hline
{\rm Our \ Output} & 0.0842 \\ \hline
{\rm Double\ Chooz} & 0.086 \pm 0.041 ({\rm stat}) \pm 0.030 ({\rm syst})
   \\ \hline
{\rm Daya\ Bay} & 0.092 \pm 0.016 ({\rm stat}) \pm 0.005 ({\rm syst}) 
   \\ \hline
{\rm RENO} & 0.113 \pm 0.013 ({\rm stat}) \pm 0.019 ({\rm syst}) 
   \\ \hline
{\rm Combined\ Fit\ (Blondel)} & 0.084 \pm 0.014 \\ \hline \hline
\end{array}
\end{eqnarray*}
\caption{Output value for $\sin^2 2 \theta_{13}$ compared with
Double Chooz \cite{DCHOOZ}, Daya Bay \cite{DayaBay}, RENO \cite{RENO} 
and the combined fit of the first two experiments together with
T2K \cite{T2K} and MINOS \cite{MINOS} by A. Blondel \cite{Blondel}.}
\label{cftherest}
\end{table} 

That our output value for $\theta_{13}$ should agree so well 
with experiment is in a sense a little fortuitous for the 
following reason.  As explained in the preceding section, of 
the 6 parameters for the present R2M2 fit, 5 are anchored 
on quantities which have been measured with fair accuracy, 
leaving only $m_{\nu_3}^D$, the Dirac mass of the heaviest 
neutrino poorly determined, on which the value of $\theta_{13}$ 
mostly depends.  The value of $m^D_{\nu_3}$ does not affect 
much the other quantities under study apart from
those associated with neutrinos.  And of the elements
of the PMNS matrix, it is the entries of the last column
which it will affect the most, as can be seen as follows.
The 3 elements of the last column may be thought of as the
direction cosines subtended by the state vector for $\nu_3$
on the charged lepton triad, the squares of which add up of
course to unity.  If the trajectory of $\balpha$ were to lie
on a plane, then $\theta_{13}$ (i.e. $\theta_{e3}$) would be
zero, so that the departure of $\theta_{13}$ from zero can be 
thought of as a measure of the nonplanarity of the trajectory. 
The further the state vector for $\nu_3$ deviates from 
the plane containing $\balpha(\mu = m_\tau)$ and $\balpha(\mu
= m_\mu)$, the more $\sin \theta_{13} = |U_{e3}| = \sqrt{1 -
|U_{\mu 3}|^2 - |U_{\tau 3}|^2}$ will differ from zero.  Now 
we have seen already in the preceding section that in view of 
the largish values of both the Cabbibo angle $|V_{us}| \sim
|V_{cd}|$ and the solar neutrino angle $\sin \theta_{12} \sim
|U_{e2}|$, the trajectory must be nonplanar to some fixed 
extent, in which case, the further that the vector for $\nu_3$ is from that of $\tau$, the greater will be the nonplanarity and 
the larger the value of $\theta_{13}$ as a result.  Thus, by
adjusting the value of the parameter $m^D_{\nu_3}$, namely the
Dirac mass of the heaviest neutrino, and hence the proximity 
of the state vector of $\nu_3$ to that of $\tau$, one can 
change the output value of $\theta_{13}$.  It was therefore 
only by a bit of luck that by putting in (\ref{params}) for 
$m^D_{\nu_3}$ the value 20 MeV, a round number, that one has
ended up with a value for $\sin^2 2 \theta_{13}$ so close to
the experimental value.  The very good agreement obtained in
Table \ref{cfT2KMINOS} and \ref{cftherest} should thus be 
regarded not so much as a successful prediction, although it
was indeed obtained before the data in Table \ref{cftherest}
appeared, but as a test of consistency with experiment, first 
for R2M2, and secondarily for the trajectory as parametrized 
in (\ref{params}), even which, of course, is already of 
significance.

As noted, the output value of $\theta_{13}$ can be changed 
by changing the input value of the parameter $m^D_{\nu_3}$ 
without affecting the quality of the rest of the fit except
that as regards the neutrinos.  The changes induced on the 
output values of the various neutrino quantities,  however, 
will be correlated so that the output value for $\theta_{13}$,
for example, will be constrained by the experimental bounds 
on $\theta_{12}$ and $\theta_{23}$, and vice versa.  Keeping 
then the rest of (\ref{params}) the same and only varying 
$m^D_{\nu_3}$, one readily comes to the following interesting 
conclusions:
\begin{itemize}
\item $\sin^2 2\theta_{13} > 0.077$ if $\sin^2 2\theta_{12}$
and $\sin^2 2\theta_{23}$ are to remain within their experimental 
bounds quoted in Table \ref{nuos} from \cite{databook}.
\item $\sin^2 2\theta_{13} \sim 0.162 > 0.15$ if $\theta_{23}$
is to be maximal, and $\sin^2 2\theta_{12}$ is to remain within 
its experimental bounds given in Table \ref{nuos}.
\end{itemize}
In other words, both of the two conditions: (i) $\theta_{13}$ 
is vanishing, and (ii) $\theta_{23}$ is maximal, much favoured 
by theoretical models previously \cite{Scott}, would have been 
excluded by the present R2M2 fit, given just the earlier 
bounds cited in Table \ref{nuos} from \cite{databook}.  But given 
now the new data summarized in Tables \ref{cfT2KMINOS} and 
\ref{cftherest} confirming a nonzero value for $\theta_{13}$, 
the R2M2 fit would then give predictions for the amount that 
$\theta_{23}$ must depart from maximal.  Indeed, if one were 
to force $\theta_{23}$ to be maximal, then the output value of 
$\sin^2 2\theta_{13}$ will have to miss the Blondel value given 
in Table \ref{cftherest} by more than $5.5 \sigma$.  Conversely, 
inputting the Blondel value for $\sin^2 2\theta_{13}$ with his 
errors, then the present fit would predict the following value 
for $\theta_{23}$:
\begin{equation}
\sin^2 2 \theta_{23} \sim 0.935 \pm 0.021,
\label{nonmax23}
\end{equation}  
a predicted departure from maximal mixing accessible to future 
tests by experiment.

\section{Concluding Remarks}

From the results summarized in sections 1 and 3, it would seem
that the R2M2 scheme is capable of a comprehensive description 
of the mass and mixing pattern of both quarks and leptons, 
which is otherwise so bewildering.  Indeed, when supplemented 
by a parametrization of the rotation trajectory, it seems to 
have succeeded even in reproducing satisfactorily the mass and 
mixing parameters.  To our knowledge, this is the first time 
that such has been achieved by any scheme.  Given that the 
generations puzzle, of which these patterns are a part, has 
long been regarded as one of the great mysteries of particle 
physics, its explanation would seem a nontrivial achievement. 
That is, of course, assuming that the results and the method 
by which they are derived manage to survive the community's 
close scrutiny.

It is anticipated that much of the scrutiny will be focussed
on the unusual property of R2M2 mentioned in the 
introduction of being able to generate nonzero masses for all 
quarks and leptons from a universal fermion mass matrix of 
rank one.  As pointed out there,
this may seem counter-intuitive in the context of a non-rotating mass
matrix and has been the 
cause of some doubts and criticisms we have encountered in 
private discussions.   As far as we 
are aware, these have not appeared in the public domain,
although the R2M2 scheme has existed for more 
than a decade and been developed in numerous publications 
since it was proposed.  But now, in view of its quite alluring consequences 
as detailed above, it would be profitable for any criticism 
to be aired and debated in public, for only when they are will the
point be properly settled.   It is after all just
a matter of logic, or so at least it seems to us, which is
capable of being sorted out.

However, even if the R2M2 scheme manages to pass all scrutiny 
on its logical validity and all its attractive results as exhibited 
in this paper are confirmed, it still leaves the question, of 
course, why the fermion mass matrix should be of rank one, and
why it should rotate.  In other words, even if it is successful
in explaining mixing and mass hierarchy, the R2M2 as it stands
is at best an empirical rule, playing the role for quarks
and leptons, perhaps, as the Mendeleev periodic table played 
for atoms or the Rydberg formula for the hydrogen spectrum.  
There is still missing the equivalent here of the quantum 
theory in atomic physics which underlies it all.  

For this, of course, one will have to go beyond the standard 
model, which leaves therefore the field wide open.  We have 
ourselves suggested a possible approach based on what we call 
framed gauge theory (FGT) \cite{efgt}, namely a gauge theory
in which are introduced as dynamical variables not only the 
usual bosonic gauge and fermionic matter fields, but also, in 
analogy to the vierbeins in gravity, the frame vectors in 
internal symmetry space as (Lorentz) scalar fields.  This idea
has the attractive feature of giving a geometric meaning not 
only to exactly 3 fermion generations but also to the standard 
Higgs field needed to break the electroweak symmetry.  It has 
been shown \cite{efgt,dfsm} also to lead to a fermion mass 
matrix which is universal, of  rank one, and rotates with 
changing scales, quite as needed here, although it is not known 
yet whether it will give a trajectory of the appropriate shape.  
Whatever its future, however, this framed standard model (FSM) 
is but one possible implementation of the R2M2 idea which has 
led to the above results.  It is not hard to imagine many other 
possibilities, and some, we believe, are already on the way 
\cite{Bjorken}.

Optimistically then, the R2M2 scheme may play the role of, 
say, the Mendeleev periodic table to particle physics.  And, 
as the periodic table has led to the quantum theory of atoms, 
so may, let us hope, the  R2M2 scheme, especially now that one
has gained here a rough idea of what the rotation trajectory
should look like, give us a hint at last to solving the great 
particle physics mystery of fermion generations.

\vspace{1cm}

\noindent \textit{Postscriptum}  

After this paper was completed, an article
appeared on the preprint arXiv by G.L. Fogli et al. \cite{Fogli} reporting
on a global analysis of neutrino mixing data.  The values they give on the mixing angles, corresponding at $1\sigma$ to:
(1) $\sin^2 2 \theta_{12} = 0.851 \pm 0.026$, (2) $\sin^2 2 \theta_{13} = 0.096^{+0.013}_{-0.012}$,
(3) $\sin^2 2 \theta_{23} = 0.958^{+0.022}_{-0.024}$,
are all consistent with our output values listed in Tables 2 and 5.  They note in particular the indication of a departure of $\theta_{23}$ from its maximal value of $\pi/4$, as was predicted
by our considerations in Section 3.  The predicted value in equation (22) obtained by inputting the Blondel value for $\theta_{13}$
\cite{Blondel} is within $1 \sigma$ of their fitted value.  However, had
we input instead their value of $\theta_{13}$ from (2) above,
we would have obtained the prediction $\sin^2 2 \theta_{23} = 0.953^{+0.016}_{-0.018}$ almost right on top of their value.
They also note an indication for a value of $\delta^\ell_{CP} =
(0.89^{+0.29}_{-0.44})\, \pi$ corresponding to $\sin \delta^\ell_{CP} =
0.34_{+0.65}^{-0.88}$ which is again consistent with the bound predicted in equation (27).

\vspace{5mm}

\appendix
\begin{flushleft}
\bf\large
{Appendix A: Some computational details to facilitate the 
verification of the results obtained above.}
\end{flushleft}

The results of the present R2M2 fit, as summarized in Sections 
1 and 3, are obtained from numerical calculations. 
In this appendix we aim to supply interested readers with 
sufficient computational details so that they can readily 
verify the results for themselves.  In the numerical results below we provide more significant figures than necessary to avoid cumulative rounding errors.  

First, we give the values of the vector $\balpha$ at the mass 
scales of the various quarks and lepton states:
\begin{eqnarray}
\balpha^\dagger(\mu = m_t) & = & (1.00000,	0.00000,	0.00000) \nonumber \\
\balpha^\dagger(\mu = m_b) & = & (0.99912	,-0.04176,	-0.00345) \nonumber \\
\balpha^\dagger(\mu = m_\tau) & = & (0.99762,	-0.06836,	-0.00880) \nonumber \\
\balpha^\dagger(\mu = m_{\nu_3}) & = & (0.74599,	-0.52127,	-0.41446) \nonumber \\
\balpha^\dagger(\mu = m_c) & = & (0.99643	,-0.08350	,-0.01281) \nonumber \\
\balpha^\dagger(\mu = m_s) & = & (0.97052,	-0.22709,	-0.08076) \nonumber \\
\balpha^\dagger(\mu = m_\mu) & = & (0.95172	,-0.28232,	-0.12048) \nonumber \\
\balpha^\dagger(\mu = m_{\nu_2}) & = & (0.27858,	-0.56685	,-0.77529) \nonumber \\
\balpha^\dagger(\mu = m_u) & = & (-0.63101	,0.76714,	0.11543) \nonumber \\
\balpha^\dagger(\mu = m_d) & = & (-0.45968,	0.81729	,0.34748) \nonumber \\
\balpha^\dagger(\mu = m_e) & = & (-0.37336,	0.81904	,0.43563) \nonumber \\
\balpha^\dagger(\mu = m_{\nu_1}) & = & (-0.71656,	0.21263,	-0.66433),
\label{balpha}
\end{eqnarray}
where we note that for consistency, the mass values of the
various states are taken as either the values listed in Table
\ref{M3} or the output values listed in Table \ref{massout}.  
From (\ref{balpha}), it can readily be checked, for example by evaluating 
their polar co-ordinates, that these vectors do indeed lie on 
the trajectory shown in Figure \ref{Muisphere}.

Secondly, we give the state vectors of the various quark and 
lepton states:
\begin{eqnarray}
{\bf t}^\dagger & = & (1.00000,0.00000,0.00000) \nonumber \\
{\bf b}^\dagger & = & (0.99912,	-0.04176,	-0.00345) \nonumber \\
\btau ^\dagger  & = & (0.99762,	-0.06836,	-0.00880) \nonumber \\
\bnu_3^\dagger  & = & (0.74599,	-0.52127,	-0.41446) \nonumber \\
{\bf c}^\dagger & = & (0.00000,	0.98843,	0.15168) \nonumber \\
{\bf s}^\dagger & = & (0.03989,	0.92267,	0.38351) \nonumber \\
\bmu^\dagger    & = & (0.06476,	0.88601,	0.45913) \nonumber \\
\bnu_2^\dagger  & = & (0.59501	,0.24219	,0.76636) \nonumber \\
{\bf u}^\dagger & = & (0.00000	,-0.15168	,0.98843) \nonumber \\
{\bf d}^\dagger & = & (-0.01284	,-0.38331	,0.92353) \nonumber \\
{\bf e}^\dagger & = & (-0.02359	,-0.45861	,0.88833) \nonumber \\
\bnu_1^\dagger  & = & (-0.29910,	-0.81831,	0.49083)
\label{statevecs}
\end{eqnarray}
It can again be readily checked that these satisfy their 
defining conditions in (\ref{tcutriad}) with respect to the
$\balpha$'s in (\ref{balpha}).

From (\ref{statevecs}), one can then easily evaluate:
\begin{itemize}
\item by taking their inner products with the corresponding 
$\balpha$'s in (\ref{balpha}), the masses of the lower generation 
states according to the formulae given in (\ref{hiermass}),
\item by taking the inner products between up and down states
the UD matrix (\ref{VUD}) for quarks and the PMNS matrix for 
leptons.
\end{itemize}

To evaluate the CKM matrix for quarks, one has further to 
supply, thirdly, the tangent vector $\btau$ and the normal
vector $\bnu$ to the trajectory at respectively $\mu = m_t$ and 
$\mu = m_b$, thus:
\begin{eqnarray}
\btau^\dagger(\mu = m_t) & = & (0.00000, 1.00000, 0.00000) \nonumber \\
\bnu^\dagger(\mu = m_t)  & = & (0.00000, 0.00000, 1.00000) \nonumber \\
\btau^\dagger(\mu = m_b) & = & (0.04179,0.98683,0.15629) \nonumber \\
\bnu^\dagger(\mu = m_b)  & = & (-0.00313,-0.15630,0.98770)
\label{taunu}
\end{eqnarray}
from which one can easily check that:
\begin{equation}
\sin \omega_U = 0.15168; \ \ \ \sin \omega_D = 0.23446.
\label{omegaUD}
\end{equation}
This then allows one to calculate the individual CKM matrix
elements, still just according to (\ref{VUD}), although, the 
vectors being now complex, the matrix elements will also be
complex.  The Jarlskog invariant will then be nonzero and can
be calculated using the formula for it given in \cite{Jarlskog,Dalitz} (see also footnote \ref{Jarlfn}).

In other words, with the information supplied above in this 
appendix, the interested reader can easily make spot checks on the cited numerical results,
or even, with some patience, verify for themselves all the
numerical results cited, and hence to satisfy themselves 
that the trajectory shown in Figure \ref{Muisphere} does indeed 
lead, via the R2M2 mechanism, to the cited mass and mixing 
patterns, in close agreement with experiment, as claimed.  And 
this, of course, is the primary question of our main concern.

As to the secondary question of whether the parametrization
(\ref{sonmu}) and (\ref{kgons}), with the values of the
parameters given in (\ref{params}), does indeed give rise to 
the trajectory shown in Figure \ref{Muisphere}, we know no other
way of checking it at present than by numerically integrating 
the Serret--Frenet--Darboux formula in (\ref{SFD2}), which 
is also straightforward to perform. 

\vspace{5mm}

\appendix
\begin{flushleft}
{\bf\large
Appendix B: The case for a PMNS matrix with a nonvanishing
CP-violating phase ($\sin\delta^\ell_{CP} \neq 0$).}
\end{flushleft}

As shown in \cite{atof2cps}, because of the presence of zero 
modes to the mass matrix at every scale, the R2M2 scheme will
always allow a chiral transformation to be made on the state 
in the $\bnu$ direction, whether of quarks or of leptons, 
without making the mass matrix complex.  The effect of this
transformation will then get transmitted by rotation to the 
mixing matrix and give it a CP-violating phase.  Indeed, in 
the present set-up where one starts with a real $\balpha$ in 
(\ref{mfact}), this seems the only way to make the otherwise 
real mixing matrix complex and to give the CP-violating phase 
a nonzero value.  What distinguishes quarks from leptons so 
far is just that such a chiral transformation on quarks will 
affect the theta-angle term in the QCD action so that this 
chiral transformation becomes now not only possible but also
necessary and in a sense unique, given that without it, and 
without having the chiral angle matched to $\theta$, strong 
interactions will be CP-violating, in contradiction to experiment.  
For leptons, on the other hand, there seems a priori no reason
as yet why such a chiral transformation should be made, and if 
it is made, no restriction on what value the chiral angle 
should take.  Hence, the suggested simplest solution in the 
text when no such chiral transformation is made at all, so the mixing matrix remains real, with 
$\sin\delta^\ell_{CP} = 0$.

There is, however, nothing in principle to stop us making a
chiral transformation on the leptons and give the PMNS matrix
a CP-violating phase.  Indeed, we shall present later in this 
appendix even some theoretical arguments for a situation where 
a chiral transformation similar to that for quarks becomes 
necessary also for leptons.  Let us first examine, however,
what effect such a chiral transformation will have on our
previous results and what sort of values it can give for 
$\delta^\ell_{CP}$ in the leptonic case.

As was found for the quark mixing matrix earlier, one effect
of the chiral transformation in the R2M2 scheme is to enhance
the (absolute) value of the top-right corner element in the
mixing matrix, namely, for the leptonic case $\sin \theta_{13}$.
Given that this quantity is tightly constrained both by our 
fit and by the new data now available, as explained at the
end of section 3, there will not be very much room for this
enhancement to operate.  Indeed, it is not hard by playing
with the parameters to ascertain that for $\theta_{13}$ to 
remain within the experimental bounds of \cite{Blondel} and
$\theta_{12}$ and $\theta_{23}$ to remain within the bounds
cited in Table \ref{nuos} from \cite{databook}, the present 
setup (which neglects possible Majorana phases) can admit only a 
chiral transformation on leptons with 
a value for $\theta$ of at most $0.64$, i.e. about half its 
previous value required for quarks as given in (\ref{params}).  
This corresponds to a CP-violating phase:
\begin{equation}
|\sin \delta^\ell_{CP}| \leq 0.31.
\label{deltaCPl}
\end{equation}
The imposition of a chiral transformation on leptons with 
a $\theta$ of the same value ($\sim 1.28$) as suggested 
in (\ref{params}) above for quarks would lead to an
output value for $\sin^2 2 \theta_{13}$ larger than the
Blondel value of Table \ref{cftherest} by nearly $4\sigma$.
The conclusion (\ref{deltaCPl}) is of interest to experiment,
but it is not easy to unravel at this stage how much of it
is due to the R2M2 scheme and how much to the particular 
manner by which it is here parametrized.

In the rest of this appendix we shall consider a theoretical
situation where a chiral transformation similar to that for 
quarks becomes necessary also for leptons.   This arises from 
the assumption in R2M2 that the vector $\balpha$ is ``universal'', 
i.e. common not only to the up and down state of quarks and 
leptons separately but common to quarks and leptons as well.  
Notice that this last assumption was not made in the original 
scheme of \cite{Fritsch,Harari} previously cited, there being 
then no real initiative for doing so.  It was made only in the 
R2M2 scheme based  partly on the observation that it seems 
to work as seen above, and partly on theoretical prejudices 
\cite{efgt,dfsm}.  Nevertheless, once it is accepted, then a 
new situation can arise, as follows.

The electroweak theory is usually interpreted as one in which
the original local gauge symmetry $su(2) \times u(1)$ is 
broken spontaneously by the Higgs field.  However, it could 
equally be interpreted, according to 't~Hooft \cite{tHooft} 
and to Banks and Rabinovici \cite{Bankovici}, as a theory in 
which the local $su(2)$ symmetry confines, what is broken 
being only a global, say $\widetilde{su}(2)$, symmetry 
associated or ``dual'' to it.  In present applications of 
the theory, the two interpretations are mathematically 
equivalent \cite{tHooft}, but some may find one intuitively
more appealing than the other. 

In the second interpretation where $su(2)$ confines, hereafter 
referred to as the confinement picture, (left-handed) quarks 
and leptons, as well as the $W-Z$ and Higgs bosons, appear not 
as fundamental fields but as bound (confined) states of such 
under $su(2)$ confinement.  For example, a left-handed lepton would 
appear as a bound state of the fundamental $su(2)$ doublet 
scalar field $\phi$ with a fundamental $su(2)$ doublet left-handed
fermion field $\psi_L$, forming together an $su(2)$ singlet 
state as confinement requires, thus:
\begin{equation}
\chi_L = \phi^\dagger \psi_L,
\label{bchi}
\end{equation}
while a left-handed quark would appear as a bound state of the same 
$\bphi$ field but now with a fundamental $su(2)$ doublet left-handed 
fermion field, say $\psi_{La}$, carrying colour $a = 1, 2, 3$, 
thus:
\begin{equation}
\chi_{La} = \phi^\dagger \psi_{La}.
\label{bchia}
\end{equation}
We notice, however, that the right-handed fundamental fermion
fields, say $\psi_R$ and $\psi_{Ra}, a = 1, 2, 3$, being by
definition $su(2)$ singlets, need not be so confined, and can 
thus already function as the right-handed leptons and quarks
respectively. 

Adopting the above confinement picture for the R2M2 scenario
under consideration, we would obtain, for the 3 basis states
$\balpha, \btau, \bnu$ appearing in (\ref{SFD2}) above, the
following left-handed leptons:
\begin{equation}
\phi^\dagger \balpha\cdot\bpsi_L, \ \ \phi^\dagger \btau\cdot\bpsi_L,
  \ \ \phi^\dagger \bnu\cdot\bpsi_L,
\label{basisleptons}
\end{equation}
and the following left-handed quarks:
\begin{equation}
\phi^\dagger \balpha\cdot\bpsi_{La}, \ \ \phi^\dagger \btau\cdot\bpsi_{La},
  \ \ \phi^\dagger \bnu\cdot\bpsi_{La},
\label{basisquarks}
\end{equation}
being both bound states, under $su(2)$ confinement, of the 
fundamental scalar $\phi$ with the 3-component (in generation space)
fundamental fermion fields $\bpsi$.

The question then arises where the new factors, $\balpha, \btau, 
\bnu$ which distinguish generations, originate, i.e. whether 
from the bound states' scalar or fermionic constituent.  Given 
that $\balpha$ which appears in the mass matrix (\ref{mfact}) 
is supposed in R2M2 to be ``universal'', i.e. the same for 
(\ref{basisleptons}) and for (\ref{basisquarks}), it seems 
this factor $\balpha$ can come only from the fundamental scalar 
field.{\footnote{We note that in the framed standard model FSM 
\cite{efgt,dfsm} mentioned earlier as providing a possible 
theoretical basis for R2M2, the factor $\balpha$ in the fermion 
mass matrix (\ref{mfact}) does indeed originate from the scalar 
constituent, as stipulated.  These scalar fields play in FSM a 
geometric role, namely that of framed vectors or vierbeins in the 
internal symmetry space.}  And since the other factors $\btau$ 
and $\bnu$ are but constructs from $\balpha$, the same conclusion 
would seem to apply to them as well.

Supposing this is true, then we have the following interesting
result.  We have noted before that a chiral transformation is to
be performed on the $\bnu$ component of the quark fields so as
to eliminate the theta-angle term \cite{atof2cps}.  We note
now that, in the confinement picture of 't~Hooft and others as
exhibited in (\ref{basisquarks}), the same chiral transform can 
be obtained just by a phase change in the $\bnu$ component of 
the scalar field.  That this is so can be seen as follows.  A 
chiral transformation $\exp(i \alpha \gamma_5)$ applied on a
quark field $\psi$ changes the left-handed component by the 
phase $- \alpha$ but the right-handed component by the opposite 
phase $\alpha$, thus:
\begin{eqnarray}
\exp(i \alpha \gamma_5) \psi & = & \exp(i \alpha \gamma_5) 
   [\half (1 - \gamma_5) \psi + \half (1 + \gamma_5) \psi]
   \nonumber \\
   & = & [\exp(-i \alpha) \half (1 - \gamma_5) \psi
          + \exp(i \alpha) \half (1 + \gamma_5) \psi]
   \nonumber \\
   & = & \exp(i \alpha) 
         [\exp(-2i \alpha) \half (1 - \gamma_5) \psi
          + \half (1 + \gamma_5) \psi].
\label{chitrans}
\end{eqnarray} 
Or equivalently, it changes the phase of the whole of $\psi$
by $\alpha$, but the left-handed component by an extra phase
$-2 \alpha$.  And since a change of the overall phase of the
quark field $\psi$ has no physical significance, this is again
equivalent to just changing the left-handed component by a
phase $-2 \alpha$ while leaving the right-handed component
unchanged.  But in the confinement picture, we see that this 
is exactly what changing the phase of the $\bnu$ component 
of $\phi$ by $-2 \alpha$ would do.  We recall there that the 
left-handed quark is a bound state of the scalar field with 
a fundamental fermion, so that a change in phase in the former 
will be transmitted to the left-handed quarks.  On the other 
hand, as noted above, the right-handed quarks are not bound 
states at all and are not affected by the change in phase of 
the scalar field.  In other words, this means that, in the 
confinement picture we have adopted, just a change in phase 
of the $\bnu$ component of $\phi$ by $-2 \alpha = \theta$ 
will eliminate the theta-angle term from the QCD 
action, and give a CP-violating phase to the CKM matrix.  
We note also that a mere change in phase of $\bnu$ will not 
affect the orthonormality of the basis Darboux triad.

The really interesting thing now is the following.  If the
chiral transformation to eliminate the theta-angle term is
indeed to be transferred to a change of phase in the $\bnu$
component of the scalar field as above suggested, then the
same chiral transformation would have to be performed also
on the lepton fields, for according to (\ref{basisleptons})
above, the left-handed leptons, in the confinement picture,
are bound states of the same scalar fields as in the quarks,
although now to some different fundamental fermion field.  As
a result, the same argument that was applied to quarks via the
rotation mechanism to deduce a CP-violating phase in the CKM
matrix will apply also to the leptons to give a CP-violating 
phase in the PMNS matrix.

However, that the ``same'' chiral transformation be applied 
to both quarks and leptons does not necessarily mean that the
value of $\theta$ is the same for both, for the two mixing 
matrices for quarks and leptons refer to very different scales
and the angle $\theta$ may itself be scale-dependent.  The
conclusion (\ref{deltaCPl}) reached above would thus seem to 
suggest, that if this theoretical scenario were to hold, then
the strong CP angle $\theta$ would have to lose about half 
its value in running from the quark to the leptonic scale.   

\vspace{5mm}

\appendix
\begin{flushleft}
{\bf\large
Appendix C: Solutions for the lightest generation mass}
\end{flushleft}

Empirically, the lightest generation is a little anomalous.
Whereas for the 2 heavier generations, one has $m_t \gg m_b
> m_\tau$ and $m_c \gg m_s \stackrel{?}{>} m_\mu$, for the
lightest generation one has $m_d > m_u > m_e$ and all of
roughly the same MeV order.  Of course, so long as one has
no explanation at all for the mass spectrum, the anomaly
is just a curiosity.  But as soon as one aims to explain
why the masses should take the values they do, then this
departure of the lightest generation from the pattern of
earlier generations would have to be accounted for.  At
first sight, the R2M2 scheme would seem to have difficulty 
with this, for in the mass matrix (\ref{mfact}) it is the 
heaviest generation which sets the scale $m_T$ for the 
masses in each species.  And if the heaviest generation
follows the above pattern, so should apparently the lower
generations by leakage from the heaviest.  This holds for
the second heaviest generation, so why the anomaly in the
lightest generation?  As indicated in Section 2, point (c),
the occurrence of multiple solutions to the mass equation
(\ref{hiermass}) give us at least a partial answer.

But how do these multiple solutions to the mass equation
(\ref{hiermass}) come about for the lightest generation, 
and especially why only for it, not for the middle one as
well?  To see this, we recall first again how solutions
to the $u$ mass come about.  The state vector ${\bf u}$
for $u$ is by definition orthogonal to the $\bt\bc$-plane.
At $\mu = m_c$, $\balpha(\mu)$ lies on the $\bt\bc$-plane and
is therefore orthogonal to ${\bf u}$ and $\langle {\bf u}|
\balpha(\mu) \rangle$ vanishes.  As $\mu$ decreases, the
vector $\balpha(\mu)$ rotates away from the $\bt\bc$-plane
giving then a nonvanishing component $\langle {\bf u}|
\balpha(\mu) \rangle$ in the ${\bf u}$ direction, which
gradually increases until at some value of $\mu$ it matches
the value $\sqrt{\mu/m_t}$ and one has a solution to the
equation (\ref{hiermass}) for the mass of $u$.  This is
then our first solution to the $u$ mass, which would occur
on the trajectory not so far from $c$, and has in actual 
fact a value of about $300 {\rm MeV}$, which is quite 
different from what experiment indicates.

However, as $\mu$ lowers further, especially when the arc
length $s$ is growing exponentially with $\mu$ as suggested
by (\ref{sonmu}), it can easily happen that the trajectory
should hit the $\bt\bc$-plane again.  Then at this point, one
has again $\langle {\bf u}|\balpha(\mu) \rangle$ vanishing,
and the same opportunity as before will present itself for
again a solution for the $u$ mass.  This is exactly what
happens for the trajectory of Figure \ref{Muisphere} and
accounts for the output mass value for $u$ listed in Table
\ref{massout}.  One sees therefore that it is rather easy
for the $u$ mass to acquire this lower solution; all it 
needs is for the trajectory to hit the $\bt\bc$-plane again at
some lower scale $\mu$, which it can easily do so long as
the trajectory is not prematurely terminated, say by a
rotational fixed point, for example, before it has gone
that far.  

The same argument would hold also for $d$ and $e$ if one
just replaces the $\bt\bc$-plane above by respectively the $\bb\bs$-
and $\btau\bmu$-plane.  These planes are defined at the 
higher scales of the two heavier generations where the
trajectory is moving relatively slowly so that the planes
are not too different from each other in terms of the arc
length $s$, nor from the $\bt\bc$-plane for $u$.  Hence the
lower solutions for $u, d, e$ would occur consecutively
and fairly close to one another on the trajectory, as is
indeed seen in Figure \ref{Muisphere}.  And since at these
lower scales, the trajectory is moving fast, meaning that
changing $s$ requires very little change in $\mu$, the
solutions for the masses of $u, d, e$ would be rather 
close in values to one another, as is indeed seen in Table
\ref{massout}.  So much then for the properties noted in
the text for the output masses for the lightest generation.  

A natural question to ask at this point is, of course, what
about the middle generation; will there not be other lower
solutions to their masses too?  Interestingly, the answer
is no, there being a rather subtle difference in the R2M2
scheme between the 2 lighter generations which makes the
likelihood of lower solution practically zero for the middle
generation.  This comes about as follows.  We recall, as we
have done already several times, that all the state vectors
are determined in the R2M2 scheme by the scale of the middle
generation.  This means in particular that by the $u$ mass
scale where one looks for solution to the mass equation
(\ref{hiermass}) for $m_u$, the state vector ${\bf u}$ is
already known, namely in that direction orthogonal to the 
$\bt\bc$-plane.  In contrast, in seeking a solution to the $c$
mass at scale $\mu = m_c$, one has first to ascertain what
is the state vector ${\bf c}$ lying on the plane orthogonal
to the state vector ${\bf t}$ of $t$.  In words, what the 
equations (\ref{tcutriad}) and (\ref{hiermass}) say is that 
one should project the vector $\balpha(\mu)$ at $\mu = m_c$
on to the plane orthogonal to ${\bf t}$.  Then the direction
of the shadow cast by $\balpha(m_c)$ on to that plane is the
vector ${\bf c}$, and the length of that shadow squared times
$m_t$ is the mass $m_c$.  In other words, for a solution to
the $c$ mass at a scale $\mu$, one will need the shadow to
be of order only $\sqrt{\mu/m_t}$, a small number especially
when $\mu$ itself is small when looking for a lower solution.
This means that for a lower solution, one will need to have
$\balpha(\mu)$ nearly parallel or antiparallel to ${\bf t}$
itself.  In other words, for $m_c$ to have a lower solution,
one will need the trajectory to pass again at a lower scale
very close to the starting point ${\bf t} = (1, 0, 0)$ or
to its antipodes $(-1, 0, 0)$, which can happen, of course, 
but is very unlikely, and it did not happen for our special
trajectory in Figure \ref{Muisphere}.  The demand is very much
more stringent than for a lower solution for the $u$ mass
which needs only that the trajectory hits again a plane, i.e.
the $\bt\bc$-plane, which it will almost always do so long as it
is not terminated prematurely.  The difference comes about
just because there are 3 generations in nature and generation
space is 3-dimensional.

We note that the trajectory in Figure \ref{Muisphere} actually
passes through the $\bt\bc$-plane at $\mu$ around 1 MeV so that
there are two solutions very close to each other to the mass 
equation (\ref{hiermass}) for the $u$ mass, one lying just
above and one just below the $\bt\bc$-plane.  Numerically, they
differ by only about 2.5 KeV, and to the accuracy one is
interested in at present, may be regarded practically as the 
same solution.  Theoretically, however, we are yet unclear
what this double solution means.  We note also that if we
further extrapolate our trajectory, it might pass through
the $\bt\bc$-plane again and give further and even lower solutions
to (\ref{hiermass}) for the $u$ mass.  We are reluctant to
do so, however, for we do not believe that the exponential
growth in rotation speed as parametrized by (\ref{sonmu})
can go on forever.  We think it is more likely that the
trajectory will eventually terminate at some low scale fixed
point as it actually does in the theoretical model for R2M2
\cite{efgt,dfsm} that we are studying.

\end{document}